\documentclass{article}

\usepackage{microtype}
\usepackage{graphicx}
\usepackage{enumitem}
\setlist{nolistsep,leftmargin=*}
\usepackage{subfigure}
\usepackage{tabularx}
\usepackage{booktabs} 

\usepackage{hyperref}

\usepackage[accepted]{tccmlicml2021}

\icmltitlerunning{Forecasting Sea Ice Concentrations using Attention-based Ensemble LSTM}

\begin{document}

\twocolumn[
\icmltitle{Sea Ice Forecasting using Attention-based Ensemble LSTM}




\icmlsetsymbol{equal}{*}

\begin{icmlauthorlist}
\icmlauthor{Sahara Ali}{to}
\icmlauthor{Yiyi Huang}{to}
\icmlauthor{Xin Huang}{to}
\icmlauthor{Jianwu Wang}{to}
\end{icmlauthorlist}

\icmlaffiliation{to}{Department of Information Systems, University of Maryland, Baltimore County, United States}

\icmlcorrespondingauthor{Jianwu Wang}{jianwu@umbc.edu}

\icmlkeywords{Ensemble LSTM, Sea Ice Prediction, Arctic Amplification, Deep Learning, ICML}

\vskip 0.3in
]



\printAffiliationsAndNotice{} 

\begin{abstract}
Accurately forecasting Arctic sea ice from sub-seasonal to seasonal scales has been a major scientific effort with fundamental challenges at play. In addition to physics-based earth system models, researchers have been applying multiple statistical and machine learning models for sea ice forecasting. Looking at the potential of data-driven sea ice forecasting, we propose an attention-based Long Short Term Memory (LSTM) ensemble method to predict monthly sea ice extent up to 1 month ahead. Using daily and monthly satellite retrieved sea ice data from NSIDC and atmospheric and oceanic variables from ERA5 reanalysis product for 39 years, we show that our multi-temporal ensemble method outperforms several baseline and recently proposed deep learning models. This will substantially improve our ability in predicting future Arctic sea ice changes, which is fundamental for forecasting transporting routes, resource development, coastal erosion, threats to Arctic coastal communities and wildlife.
\end{abstract}

\section{Introduction}
\label{submission}

Over the last three decades, the warming of the sea ice in the Arctic has been almost twice faster than the rest of the world. This phenomenon is also called Arctic amplification \citep{holland2003polar}.This amplification will further alter the climate patterns beyond the Arctic region and lead to more intense and more frequent extreme weather events. A recent example of this is the 2021’s historic cold snap in Texas and Oklahoma, where more than 3000 daily low temperature records were broken, resulting in at least 176 fatalities and \$195 billion economic loss \cite{freedman_muyskens_samenow_2021}. The sea ice plays a key role in the Arctic climate system and it has been fallen by half since 1979 when the satellite observations were available \citep{serreze2015}. On current trends, the Arctic ocean could be sea ice free by 2050 \citep{notz2018trajectory}. Such rapid changes has profound local and global impacts on transporting routes, resource development, coastal erosion, military and civilian infrastructure, Arctic coastal communities (hunting and transportation by indigenous populations),  wildlife (e.g., polar bear access to food sources).

Though most of the scientists agree that this rapid Arctic warming is a sign of human-caused climate change, studying the causes of Arctic amplification and forecasting sea ice has become one of the most hyped questions in the Earth Science research \citep{holland2019changing}. Current operational sea ice forecasting systems are mainly based on coupled Earth System Models, which estimates the solution to differential equations of fluid motion and thermodynamics to obtain time and space dependent values for various variables in the atmosphere, ocean or sea ice. However, these physics-based models perform no better than simple statistical methods at lead times of two months and beyond. Over the last few years, there are many studies focusing on sea ice forecasting using data-driven Artificial Intelligence (AI) approaches like machine learning and deep learning. \citet{rs9121305} compared the Long Short-Term Memory (LSTM) model with a traditional statistical model, and they found that the LSTM showed good performance for 1-month sea ice concentration (SIC) prediction, with less than 9\% average monthly errors. However, there is lower predictability during the melting season, with the root-mean-squared-error (RMSE) of 11.09\% from July to September. \citet{rs_wang_2017} showed the superiority of the convolutional neural networks (CNN) in SIC prediction with an estimated RMSE of 22\% compared to a multi-layer preceptron model. \citet{rs_kim_2019} applied deep neural networks with Bayesian model averaging ensemble to predict the SIC in the next 10 to 20 years and found out the annual RMSE of 19.4\%. More recently, \citet{tc-14-1083-2020} built a novel 1-month SIC prediction model with CNNs using eight predictors to predict SIC both temporally and spatially. They showed that CNNs had better performance than a random forest model and the persistence model based on the monthly trend, with the RMSE of 5.76\%. \citet{jmse9030330} proposed a 1-day SIC prediction model based on the Convolutional LSTM and they concluded that the predictability of Convolutional LSTM always performed better than that of the CNN, with an RMSE of 11.24\% through 10 consecutive days of iterative prediction. 

In light of above related work, we introduce a sea ice forecasting system with Attention-based Ensemble Long Short-Term Memory (LSTM) networks to predict monthly sea ice extent with a lead time of 1 month. We have published our open-access code on  \href{https://github.com/big-data-lab-umbc/sea-ice-prediction/tree/main/climate-change-ai-workshop}{github \footnote{https://github.com/big-data-lab-umbc/sea-ice-prediction}}. The major goal of this study is to work on deep learning based forecasting models that can utilize multi-temporal data to forecast daily, weekly and monthly sea-ice extent with comparatively high accuracy. Our major contributions are as follows.
\begin{itemize}[topsep=0pt, partopsep=0pt]
    \item{We propose an ensembling method for multi-temporal Deep Learning models that performs ensembling of constituent models to give optimal sea ice predictions for next 1 month.}
    \item{We introduce an attention mechanism in our ensemble that lets model pay more or less attention to different features learned by constituent Deep Learning models leading to lower loss and higher accuracy.}
\end{itemize}
\begin{figure*}[!htbp]
    \centering
    \includegraphics[width=0.8\textwidth]{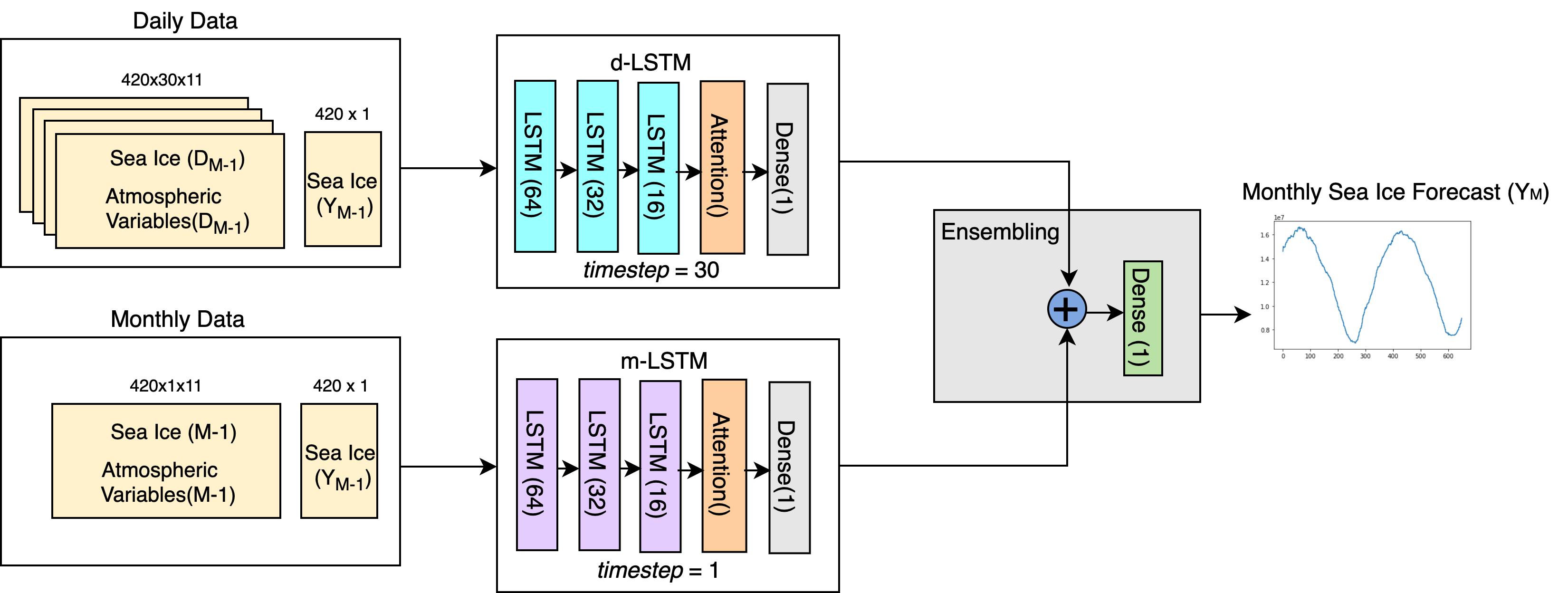}
    \caption{Overall Architecture of Attention-based LSTM Ensemble (EA-LSTM).}
    \label{architecture}
\end{figure*}
\section{Data and Method}

\subsection{Dataset}
In this study, we use ten atmospheric and ocean variables obtained from ERA-5 global reanalysis product and sea ice extent values, derived from sea ice concentrations, that we obtained from the Nimbus-7 SSMR and DMSP SSM/I-SSMIS passive microwave data version 1 \cite{cavalieri_parkinson_gloersen_zwally_1996} provided by the National Snow and Ice Data Center. These variables are enlisted in Table 1. We created two time-series combining both sea ice extent and atmospheric variables for a span of 39 years, from 1979 to 2018. In the first time-series, monthly gridded data during 1980-2018 has been averaged over the Arctic north of 25$^{\circ}$ N using area-weighted method. In the second time-series, daily gridded data has been averaged over the same spatial location.
Throughout the experiments, we use first 34 years of data for training and last 5 years of data for testing.
\begin{table}[!htbp]
\caption{Variables included in the Dataset}
\label{sample-table}
\begin{center}
\begin{small}
\vskip 0.15in
\begin{sc}
\begin{tabular}{cccc}
\toprule
Variable & Range & Unit\\
\midrule
surface pressure & [400,1100] & hPa  \\
wind velocity & [0,40] & m/s \\
specific humidity & [0,0.1] & kg/kg\\
air temperature & [200,350] & K\\
shortwave radiation & [0,1500] & $W/m^2$ \\
longwave radiation & [0,700] & $W/m^2$ \\
rain rate & [0,800] & mm/day \\
snowfall rate & [0,200] & mm/day\\
sea surface temperature & [200,350] & K  \\
sea surface salinity & [0,50] & PSU \\
sea ice concentration & [0, 100] & \% \\
\bottomrule
\end{tabular}
\end{sc}
\end{small}
\end{center}
\vskip -0.1in
\end{table}
\subsection{Preprocessing}
To feed our data to LSTM model, we reshape our $N$x$11$ two-dimensional datasets to $M$x$T$x$11$ three-dimensional datasets in such a manner that each row of predictors corresponds to (M-1)$th$ month's sea-ice value. Here N represents total number of data samples, M represents total number of months and T represents the timestep. We achieve this by removing the first row of sea-ice values and Nth record from monthly and daily data, the details of which are mentioned in sections below. In both the approaches, we rescale the data using Standard Scaler normalization technique.
\subsubsection{Daily Data}
We first needed to reshape the daily data to add a third dimension in order to get chunks of 30 X$_{M-1}$ predictors corresponding to each Y$_{M}$ predictand. For this, we first removed the 31st days from specific months. Then we augmented 28th or 29th day of February once or twice depending upon the occurrence of leap year. The final rows were perfect multiples of 30. We then reshaped the N rows of 11 predictors to Mx30x11 3D matrix and removed the last Mth row to align the dataset with $M-1$ monthly sea ice values. Here, 30 represents the timestep T for one month's data. With this arrangement, we get January 1979's data values against February 1979's sea-ice extent value and so on. 
\subsubsection{Monthly Data}
Since this version of data already corresponded to monthly sea-ice values, we reshape the Mx11 data matrix to Mx1x11 data matrix. Here 1 represents timestep T for one month. To incorporate lead time of 1, we remove last Mth record of monthly data to align it with $M-1$ sea-ice values such as January 1979's monthly record corresponds to February 1979's sea-ice value and so on.
\subsection{Overall Architecture}
We propose an attention-based ensemble method for multi-temporal LSTM modeling to predict monthly sea ice extent with a lead time of 1 month. To cater the two temporal frequencies of datasets, we first design two simple many-to-one LSTM networks, daily LSTM or d-LSTM, and monthly LSTM or m-LSTM. We then concatenate the results from these two branches with and without attention to get monthly predictions for the sea-ice values.  All three variants of models are developed using Keras Functional API with Tensorflow backend, optimized using 'Adam' optimizer and evaluated using 'mean-squared-error' loss. To avoid over-training, the models are trained using the Early Stopping evaluation criteria. The overall architecture of our attention-based ensemble method is illustrated in Figure~\ref{architecture}. Here, $D_{M-1}$ and $M-1$ corresponds to daily and monthly records from preceding month and $Y_{M-1}$ represent sea-ice values from preceding month whereas $Y_{M}$ represents predictions for the next month.
Here we are combining only two temporal models, owing to the availability of two temporal datasets. However, the proposed architecture can incorporate any number of Deep Learning models in its concatenation part; the final fully connected layer learns weights to assign to individual models and returns prediction based on those weights.

\subsection{Daily-to-Monthly Prediction Network (d-LSTM)}
We first designed a simple LSTM network with three LSTM layers, one Dropout layer and two fully connected layers. This model takes in as input a three dimensional array $N$x$T$x$P$ of daily data, where N is the batch size, T is the timestep and P is the number of predictors. In our case, this setting is equivalent to $M$x$30$x$11$. The main motivation behind developing this network was to overcome the small-data problem faced by a deep learning model that is trained on few hundreds of monthly data and predicts monthly sea-ice values. We train this model on a dataset of 12,240 records, that is 34 years. The corresponding prediction is the monthly means of sea-ice extent for ($M+1$)$th$ month. To our knowledge, combining lower temporal frequency data to predict a higher temporal frequency value is still a novel approach in data-driven atmospheric forecasting.

\subsection{Monthly-to-Monthly Prediction Network (m-LSTM)}
We then designed another simple LSTM network with the similar architecture as d-LSTM. However, instead of a timestep of 30, this models takes in a timestep of 1, representing one month. We train this model on the monthly dataset of 420 records, that is 34 years. Similar to d-LSTM, here the corresponding predictand is the monthly mean of sea-ice extent for ($M+1$)$th$ month. 

\subsection{Attention-based Ensemble Model (EA-LSTM)}

\textbf{Attention Mechanism.}
Not all hidden states of recurrent neural network  contribute equally to the predictions. Inspired by attention mechanism introduced in  \cite{luong-etal-2015-effective}, we add an intermediary self-attention layer on top of the final LSTM layers in both d-LSTM and  m-LSTM, in order to identify which hidden states contribute more to the target prediction. The attention mechanism assigns importance scores to the different hidden states of LSTM model enabling the model to focus on most relevant features within the input. Specifically, at each time step $t$, our attention layer first takes as input the hidden states $h_t$ at the top layer of a stacking LSTM, it then infers a context vector $c_t$ that captures relevant hidden information based on current target state $h_t$ and all source states $h_s$ of the LSTM model.  Attention mechanism helps improve the deep learning model by attending to relevant hidden states so that it can significantly reduces the error of the prediction. 

\textbf{Ensemble Model.}
In all recent research work conducted on sea ice predictions, we have seen models trained on same temporal frequency as their predictions. However, in our case, m-LSTM itself cannot be a good predictive method for our dataset due to its small volume. Since the monthly dataset suffers from small-data problem, we combined the two approaches using an attention-based ensemble technique. To ensemble the two temporal models, we retrieve the model output from each of the fully connected layers of both LSTM networks to get the individual model predictions. We then concatenate the predictions retrieved from both LSTM branches and add a fully connected layer to learn the concatenation weights of both the models give final monthly sea-ice predictions. 
We give a generic formulation of our ensemble in Eq. 1 where y is the final model outcome, $f_1,..f_n$ represent the $N$ constituent inner models and $w_n$ is the weight assigned to individual model. In our case, $N = 2$.
\begin{equation}
    p(y|f_1,..f_n) = \sum_{n=1}^{N}w_{n}f_{n}
\end{equation}

\section{Results \& Analysis}
We present results from the experiments conducted on baseline models as well as our proposed deep learning models. We present our analysis by comparing the results from all experiments using the percentage RMSE loss and R-squared score.
\subsection{Baseline Experiments}
To evaluate the comparative performance of our proposed model, we performed sea-ice forecasting on monthly dataset with a lead time of 1 using five renowned Machine Learning models enlisted in Table 2. All these models were trained on first 34 years of data, that is, 420 records and tested on last 5 years, that is, 80 records. The results achieved from these models are tabulated in Table 2. Among these baseline models, we see that both ensemble models Random Forest and XGBoost have the highest prediction performance on the monthly dataset.

\begin{table}[!htbp]
\caption{Experimental results from baseline (first five rows) and proposed (last four rows) models on Sea Ice prediction with a lead time of 1 month.}
\label{sample-table}
\vskip 0.15in
\begin{center}
\begin{small}
\begin{sc}
\begin{tabular}{lcccr}
\toprule
Model & $R^{2}$ Score & \%RMSE \\
\midrule
Linear Regression & 0.974 & 5.05 \\
Decision Tree & 0.963 & 6.00 \\
Random Forest & 0.978 & 4.63 \\
XGBoost & 0.976 & 4.83 \\
Polynomial Regression & 0.966 & 5.76 \\
$d-$LSTM & 0.980 & 4.45 \\
$m-$LSTM & 0.981 & 4.21 \\
E-LSTM & 0.977 & 4.60 \\
EA-LSTM & \textbf{0.982} & \textbf{4.11} \\
\bottomrule
\end{tabular}
\end{sc}
\end{small}
\end{center}
\vskip -0.1in
\end{table}
\begin{figure}[!htbp]
\vskip 0.2in
\begin{center}
\centerline{\includegraphics[width=1.2\columnwidth]{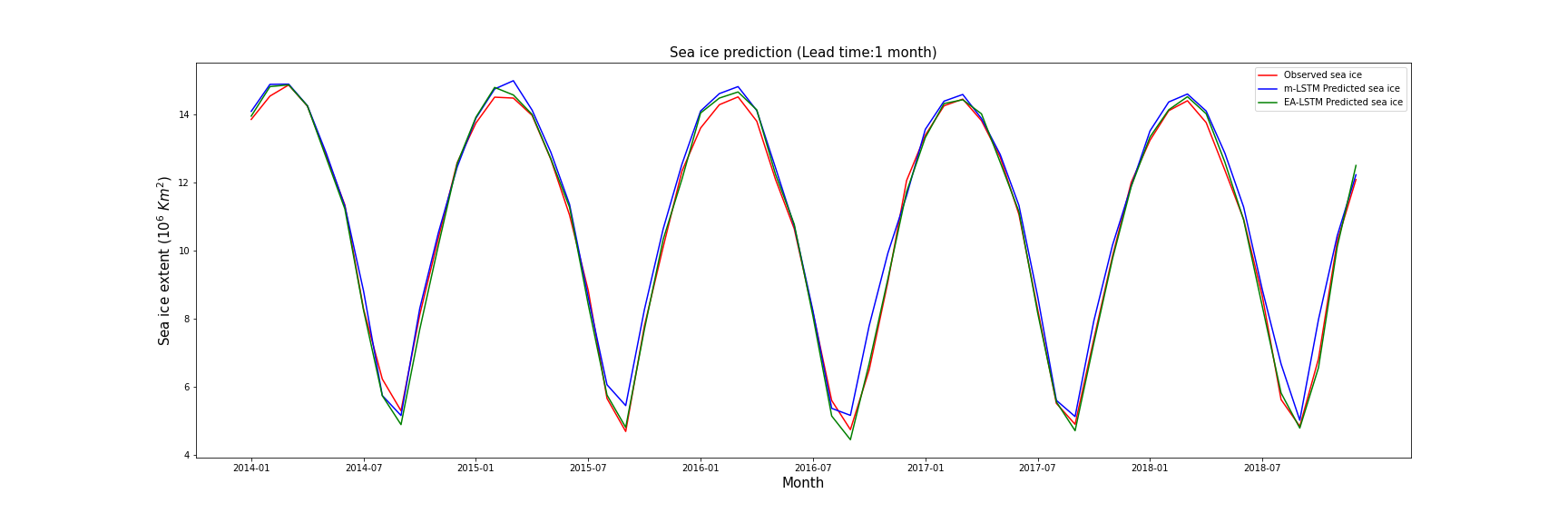}}
\caption{Time-Series Plot of Sea Ice observational data, m-LSTM predictions and EA-LSTM predictions for January 2014 - December 2018.}
\label{predicted-values}
\end{center}
\vskip -0.2in
\end{figure}
\subsection{Comparative Analysis}
We first evaluated the performance of simple d-LSTM and m-LSTM models. Through our experiment, we see that d-LSTM gives promising predictions as it is trained on daily data and predicts monthly mean sea ice values. We further observed that m-LSTM not only out-performs d-LSTM but also provides competitive results compared to multiple baseline models. We then evaluated the performance of our ensemble method with and without attention and observed that attention mechanism not only improved the overall performance but also significantly reduced the model loss. A tabulated summary of these results can be seen in Table 2.
Comparing our results with the performance of similar methods proposed by \citet{rs9121305} and \citet{tc-14-1083-2020}, we see a significant performance improvement in our sea-ice predictions with around 1.65\% and 8\% reduction in RMSE loss respectively, as compared to their results. 
\\
Figure~\ref{predicted-values} shows the temporal predictions from m-LSTM and EA-LSTM against the observed sea-ice values for the time period of January 2014 to December 2018. Looking at the observed versus predicted graphs of sea-ice, we see that our EA-LSTM method gives predictions that are more aligned with the observed values; it also gives better predictions of sea-ice extent values in the melting season of July-September as compared to simple m-LSTM, giving us more confidence in the attention-based ensemble method. 

\section{Conclusion \& Future Work}
In this paper, we present an attention-based ensemble method to predict monthly sea-ice values with a lead time of 1 month. Our proposed model can combine models having different temporal resolutions and learn their weighted average using attention-based mechanism. We started off with simple daily-to-monthly and monthly-to-monthly prediction models and moved forward to develop an ensemble model that learns from multi-temporal datasets and gives sea-ice predictions with a commendable RMSE loss of 4.11\%. Though we trained the ensemble using only two model branches, our proposed architecture is expandable to any number of model branches that can be trained on different temporal resolutions. Through our experiments, we showed how attention improves model performance and reduces overall loss. Comparing our results with baseline models and related research work, we see our attention-based ensemble technique gives promising results for forecasting monthly sea-ice extent. 

It should be noted that there are several limitations with
current study. First, the uncertainties in satellite retrieved sea ice observation and ERA-5 reanalysis product cannot be neglected, which will exert a large influence on our results. In addition, the seasonality and inter-annual variability are not well handled by the current method. In the future, we are going to apply the same method to deseasonalized time series. In the meantime, the simulations from different models will be used to better understand the impacts of inter-annual variability on sea ice predictability.

For future work, we will also extend our proposed ensemble method to combine spatio-temporal models with different spatial and temporal resolutions. We also plan to visualize the attention weights learned from these intermediary attention layers to better understand the underlying working of deep learning models and to interpret what the ensemble model learns.

The 1-month sea ice forecast is our starting point. In the future, we will go beyond the seasonal sea ice forecast and investigate how ensemble method can help us to improve the long-term sea ice projection. The sea ice coverage and length of open water season will inform future planning of military, civilian, and commercial infrastructure including buildings and marine vessels. In the meantime, we should also be aware of some potential risks with a more accurate sea ice forecast (e.g., more aggressive resource extraction in the Arctic). Nevertheless, this work will not only help us to better predict the future climate, but also promote local communities and the broader international society to respond and adapt to a changing Arctic with increased greenhouse emissions.

\section*{Acknowledgment}
This work is supported by grants CAREER: Big Data Climate Causality Analytics (OAC--1942714), REU Site:
Online Interdisciplinary Big Data Analytics in Science and Engineering (OAC--2050943), and CyberTraining: DSE: Cross-Training of Researchers in Computing, Applied Mathematics and Atmospheric Sciences using Advanced Cyberinfrastructure Resources (OAC--1730250) from the National Science Foundation.

\bibliography{workshop_paper}
\bibliographystyle{tccmlicml2021}

\end{document}